\documentclass[smallextended,referee,envcountsect]{svjour3}
\smartqed
\usepackage{graphicx,amsmath,amssymb}

\def\Mc{{\cal M}}

\def\Fc{{\cal F}}

\def \a{\alpha}

\def \k{\kappa}

\def \e{\varepsilon}
\newcommand\R{{\mathbb{R}}}
\newcommand\N{{\mathbb{N}}}

\newtheorem{assumption}{Assumption}

\begin{document}

\title{Risk-Neutral Pricing for Arbitrage Pricing Theory}

\author{Laurence Carassus \and Mikl\'os R\'asonyi}

\institute{Laurence Carassus \at
             L\'{e}onard de Vinci P\^ole Universitaire, Research Center, 92 916 Paris La D\'{e}fense, France \\ and LMR, UMR 9008, Universit\'{e} de Reims-Champagne Ardenne, France\\
              laurence.carassus@devinci.fr           \and
           Mikl\'os R\'asonyi,  Corresponding author  \at
              Alfr\'ed R\'enyi Institute of
Mathematics, Budapest, Hungary\\ rasonyi@renyi.hu}
\date{Received: date / Accepted: date}

\maketitle

\begin{abstract}
We consider infinite dimensional optimization problems motivated by the financial model called Arbitrage Pricing Theory. Using probabilistic and functional analytic tools, we provide a dual characterization  of the super-replication cost. 
Then, we show the existence of optimal strategies for investors maximizing their expected utility and the convergence of  their reservation prices  to the super-replication cost as their risk-aversion tends to infinity.
\end{abstract}



\section{Introduction}

We study infinite dimensional optimization problems motivated by a celebrated financial theory  called Arbitrage Pricing Theory (APT).
We first expose  the economic and financial background  related to APT and show how important it is for both the financial
mathematics and the mathematical economics communities.   
Then, we explain our contributions to this widely studied field 
together with their mathematical aspects.

Arbitrage Pricing Theory was originally introduced by Ross (see \cite{ross,rossr}),  and later extended by \cite{huberman,chamberlainr},
and numerous other authors.
The APT assumes an approximate factor model and states that the
risky asset returns in a ``large'' financial
market are linearly dependent on a finite set of random variables, termed factors,
in a way that the residuals are uncorrelated with the factors and with each other.
One of the desirable aspects of the APT is that it can be empirically tested as argued, for example, in \cite{dybvig-ross}.
These conclusions had a huge bearing on empirical work: see for instance
\cite{brown-weinstein}. Papers on the theoretical aspects of APT mainly focused on showing  that the model is a good approximation in a sequence of economies
when there are ``sufficiently many" assets (see for example, \cite{ross,huberman,chamberlainr}).

Ross derives the APT pricing formula  under the assumption of absence of asymptotic arbitrage in the
sense that a sequence of asymptotically costless and riskless finite portfolios cannot 
yield a positive return in the limit.
Mathematical
finance subsequently took up the idea of a market
involving a sequence of markets with an increasing number of assets in the so-called
theory of large financial markets (see, among other papers, \cite{{kabanov-kramkov},kabanov-kramkov1,klein,{iw1}}). Authors mainly  studied the characterization of
asymptotic notions of absence of arbitrage, using sequences of portfolios
involving finitely many assets, where the classical notion of no-arbitrage holds true, i.e., non-negative portfolios with zero cost should have zero return.
For the sake
of generality, continuous trading was assumed in the overwhelming majority of related papers.
But these generalizations somehow overshadowed the highly original ideas suggested in \cite{ross}, where a one-step
model was considered. They did not answer the following  natural question either: in the APT is there a way to consider strategies
involving possibly all the infinitely many assets and to exclude exact arbitrage for them rather than considering only asymptotic notions
of arbitrage?
A first answer was given in \cite{khan-sun3} in a measure-theoretical setup. Then, \cite{ijtaf,jmaa} proposed a 
straightforward concept of portfolios
using infinitely many assets, which we will use in the present paper, too:
see Section \ref{lmm} below. This notion leads to the existence of  
equivalent risk-neutral (or martingale) probability measures.

While questions of arbitrage for APT have been extensively studied by the economics and financial mathematics communities, other crucial topics -- such as utility maximization or pricing --
received little attention though these are important questions in today's markets, where there is a vast array of available assets. This is particularly conspicuous
in the credit market, where bonds of various maturities and issuers indeed constitute
an entity that may be best viewed as a large financial market (see \cite{josef}). Questions of
pricing inevitably arise and current literature on APT does not provide satisfactory answers. A standard problem is calculating the superreplication cost of a claim $G$. It is the minimal amount  needed for an agent selling $G$ in order to superreplicate  $G$  by trading in the market. This is the hedging price with no risk and, to the best of our knowledge, it was first introduced in \cite{bensaid} in the context of transaction costs.
In complete markets with finitely many assets, the superreplication cost  is just the cash flow's 
expectation computed under the unique martingale measure. When such markets are incomplete,
there exists
a so-called dual representation in terms of supremum of those expectations computed under 
each risk-neutral probability measure,
see \cite{cvitanic-karatzas} and the references in \cite{follmer-schied}.  
Our first contribution is such a representation theorem
for APT under mild conditions (see Theorem \ref{thmsur}).
The proof is based on functional analytic techniques such as the Marcinkiewicz-Zygmund inequality or 
the Banach-Sacks property.
The uniform integrability property proved in Lemma \ref{miki} together with dual methods
(using risk-neutral probabilities) allow to prove, for the first time in the context of APT,  the closure in probability of the set of attainable terminal payoffs, 
after possibly throwing away money (see Proposition \ref{L3} and Corollary \ref{Cferme}). 
We also prove a characterization of the no-arbitrage condition in a so-called 
\lq\lq{}quantitative\rq\rq{} form (see Proposition \ref{aoaquant}), which will be crucial in the rest of the paper. 
We mention \cite{paolo}, where the superhedging of contingent claims has already been considered in the general
context of continuous-time large financial markets. That paper, however, relies on the notion of generalized
portfolios, which fail to have a natural interpretation unlike the straightforward portfolio concept we use here.

Next, we consider economic agents whose preferences are of von Neumann-Morgenstern type (see \cite{von}), i.e., they
are represented by
concave increasing utility functions. In our APT framework, we are  able to prove  the existence of optimizers for such utility functions on the positive
real axis (see Theorem \ref{csonti}). Such results are standard for finitely many assets 
(see the references in  \cite{follmer-schied}),
but in the present context we face infinite-dimensional portfolios. In the setting of APT, we mention \cite{oleksi}, which relies on the notion of generalized
portfolios. 
Utility functions defined on the real line (i.e., admitting losses) have been
considered in \cite{ijtaf,jmaa} (we expose the differences between these two papers and ours in Remark \ref{diffmiklos} 
below). Our quantitative no-arbitrage characterization allows to prove a key boundedness 
condition on the set of admissible strategies (see Lemma \ref{hborne}) and 
the existence of an  optimal solution.
Finally, we establish that, when risk aversion tends to infinity,
the utility indifference (or reservation) prices (see \cite{hodges-neuberger}) tend to the superreplication price. This links in a
nice way investors' price calculations to the preference-free cost of superhedging (see Theorem \ref{main}). It also justifies
the use of a cheaper, preference-based price instead of the super-replication price, which may be 
too onerous.

The model is presented in Section \ref{lmm}. Concepts of no-arbitrage are discussed in Section \ref{noab}.
The dual characterization of superreplication prices is   given in Section \ref{supe}, the utility maximization
problem is treated in Section \ref{negu}. The asymptotics of reservation prices in the high risk-aversion
regime is investigated in Section \ref{cove} and Section \ref{conclu} concludes.


\section{The Large Market Model}\label{lmm}

Let $({\Omega}, \Fc, P)$ be a probability space. We consider a one-step economy, 
which contains a countable number of tradeable assets.
The price of asset $i\in\mathbb{N}$ is given by $(S^i_t)_{\{t \in \{0,1\}\}}$. The returns  $R_i$, $i\in\mathbb{N}$ represent the profit (or loss) created tomorrow from
investing one dollar's worth of asset $i$ today, i.e., $R_i={S^i_1}/{S^i_0}-1$.
We briefly describe below our version of the Arbitrage Pricing Model, identical to that of
\cite{kabanov-kramkov1,ijtaf,jmaa,def}, which is a special case of the model presented
in \cite{ross,huberman}. Asset $0$ represents a riskless investment and, for simplicity, we assume a 
zero rate of return, i.e.,  $R_{0}=0$. 
We assume that the other assets' returns are given by
\begin{eqnarray*}
R_i &:=& \mu_i+\bar{\beta}_i\varepsilon_i,\quad 1\leq i\leq m; \quad 
R_i := \mu_i+\sum_{j=1}^m \beta^j_i\varepsilon_j+\bar{\beta}_i
\varepsilon_i,\quad i>m,
\end{eqnarray*}
where the $\varepsilon_i$ are random variables and $\mu_i,\,\beta^j_i,\,\overline{\beta}_i$ are
constants. 
The random
variables $\varepsilon_i$, $1 \leq i \leq m$ serve
as \emph{factors}, which influence the return on all the assets $i\geq 1,$ while $\varepsilon_i,\ {i>m}$
are random sources particular to the individual assets $R_i$, $i>m$.
\begin{assumption}
\label{un}
The $\varepsilon_i$ are square-integrable, independent random variables satisfying 
$E(\varepsilon_i)=0$ and $E\left(\varepsilon_i^2\right)=1$ for all $i\geq 1.$
\end{assumption}


Assuming that
$\bar{\beta}_i\neq 0$, $i\geq 1$, we reparametrize the model using
\begin{eqnarray*}
b_i := -\frac{\mu_i}{\bar{\beta}_i},\quad 1\leq i\leq m;  &\;  \;&
b_i := -\frac{\mu_i}{\bar{\beta}_i}+
\sum_{j=1}^m \frac{\mu_j\beta^j_i}{\bar{\beta}_j\bar{\beta_i}},\quad i>m
\end{eqnarray*}
and set $b:=(b_i)_{i\geq 1}$.
Asset returns then take the following form:
\begin{eqnarray*}
R_i &=& \bar{\beta}_i(\varepsilon_i-b_i),\quad 1\leq i\leq m; \quad 
R_i = \sum_{j=1}^m\beta_i^j(\varepsilon_j-b_j)+\bar{\beta}_i(\varepsilon_i-
b_i),\quad i>m.
\end{eqnarray*}
For some $n \in\mathbb{N}$, a portfolio
$\phi$ in the assets $0,\ldots,n$ is
an arbitrary sequence $(\phi_i)_{0\leq i\leq n}$
of real numbers satisfying $\sum_{i=0}^n \phi_iS^i_0=x$, 
where $x$ is a given  initial wealth.
As $S^0_1=S^0_0$ such a portfolio will have value tomorrow given by
\begin{eqnarray*}
V^{x,\phi}_n & := & \sum_{i=0}^{n} \phi_i S^i_1
= x + \sum_{i=1}^{n} \phi_i S^i_0R_i=  x+\sum_{i=1}^n h_i (\varepsilon_i-b_i)=: V^{x,h}_n,
\end{eqnarray*}
for some $(h_1,\ldots,h_n)\in\mathbb{R}^n$, using our parametrization. \\
The value tomorrow  that can be attained using finitely many
assets is given by
$ 
J^x:=\cup_{n \geq 1} \left\{V^{x,h}_n: \, (h_1,\ldots,h_n)\in\mathbb{R}^n\right\}.
$
As $J^x$ fails to be closed in any reasonable sense, we consider
strategies, which can use infinitely many assets. This is 
desirable from an economic point of view (see \cite{khan-sun3}). 
Let 
\begin{eqnarray*}
\Phi:\; \ell_2:=\left\{(h_i)_{i\geq 1}: \, \sum_{i=1}^{\infty}h_i^2<\infty\right\} & \to & L^2(P):=\{X:\Omega \to \R, \, E|X|^2< \infty\}\\
x & \mapsto & \Phi(h):=\sum_{i=1}^{\infty}h_i\e_i. 
\end{eqnarray*}
Recall that the spaces $\ell_2$ and $L^2(P)$ are Hilbert spaces with the respectiv norm 
$||h||_{\ell_2}:=\sqrt{\sum_{i=1}^{\infty}h_i^2}$ and $||X||_{L^2(P)}:=\sqrt{E(|X|^2)}$. 
The infinite sum in $\Phi(h)$ has to be understood as the limit in $L^2(P)$ of $(\sum_{i=1}^{n}h_i\e_i)_{n \geq 1}$, which are Cauchy sequences.
Indeed, let $ h \in \ell_2$, under Assumption \ref{un}, for $p>n$,
\begin{eqnarray*}
E\left(\left(\sum_{i=1}^{p}h_i\e_i- \sum_{i=1}^{n}h_i\e_i\right)^2\right)
 & =&
\sum_{i=n+1}^{p}h_i^2\leq \sum_{i=n+1}^{\infty}h_i^2,
\end{eqnarray*}
which can be arbitrarily small for $n$ large enough.
Actually, under Assumption \ref{un}, $\Phi$ is even an isometry, 
i.e.,\ $|| \Phi(h) ||^2_{L^2(P)}=\sum_{i=1}^{\infty}h_i^2
= \|h \|^2_{\ell_2}$.
We would like to give sense (as an $L^2(P)$ limit of a sequence of finite sums) to
the portfolio value 
$V^{x,h}:=x+\sum_{i=1}^{\infty} h_i (\varepsilon_i-b_i).$
Since
\begin{eqnarray}
\label{iso}
E\left(\left(\sum_{i=1}^{p}h_i(\e_i-b_i)- \sum_{i=1}^{n}h_i (\e_i-b_i)\right)^2\right)
& =&
\sum_{i=n+1}^{p}h_i^2 +
\sum_{i=n+1}^{p}h_{i}^{2}b_i^2,
\end{eqnarray}
we need the following hypothesis.
\begin{assumption}\label{b}
We have that $b \in \ell_2$.
\end{assumption}
Then, \eqref{iso} shows that $(\sum_{i=1}^{n}h_i(\e_i-b_i))_{n \geq 1}$ is a Cauchy-sequence in $L^2(P)$
and $V^{x,h}$ is well-defined.
Notice furthermore that
\begin{equation}
\label{isol}
E\left(\left(\sum_{i=1}^{\infty}h_i(\e_i-b_i)\right)^2\right) = \sum_{i=1}^{\infty}h_i^2 +
\sum
_{i=1}^{\infty}h_i^2 b_i^2 
\leq  (1+\|b\|_{\ell_2}^2) \| h\|^2_{\ell_2}<\infty.
\end{equation}
From now on, we will use the notation 
$
\langle h, \e-b\rangle:=\sum_{i=1}^{\infty}h_i(\e_i -b_i).
$ 
Under Assumptions \ref{un} and \ref{b}, the portfolio values tomorrow  that can be attained using infinitely many
assets with a strategy in $\ell_2$  is thus given by
\begin{eqnarray*}
K^x:=\{V^{x,h}: \, h\in \ell_2\}=\{x+\langle h, \e-b\rangle: \;h\in \ell_2\}.
\end{eqnarray*}

\section{No-Arbitrage in Large Markets}\label{noab}
In Arbitrage Pricing Theory,  the classical notion of arbitrage is the asymptotic arbitrage in the sense of \cite{ross} and \cite{huberman}.
\begin{definition}
There is an asymptotic arbitrage, if there exists a sequence of strategies
$(h(n))_{n \geq 1}$, with $h(n)= (h(n)_i)_{1 \leq i \leq n}$, such that
$$E(V_n^{x,h(n)}) \underset{n\rightarrow +\infty}{\longrightarrow} \infty \mbox{ and } \mbox{Var}(V_n^{x,h(n)}) \underset{n\rightarrow +\infty}{\longrightarrow} 0.$$ 
If there exists no such sequence, then we say that there is absence of asymptotic arbitrage (AAA).
\end{definition}

We would like to understand the link between AAA and the classical definition of no-arbitrage, 
as formulated in the next definition.


\begin{definition}
\label{AOAN}
The no-arbitrage condition on a ``small market'' with $N$ random sources for some $N \geq 1$ holds true,  if 
$P\left(\sum_{i=1}^N h_i(\e_i - b_i) \geq 0\right)=1$ for $(h_1,\ldots, h_N) \in \R^N$  implies that
$h_1=\ldots= h_N=0$. This  is called AOA($N$). 
\end{definition}

We prove that under the following assumption there is absence of arbitrage in any of the small markets containing $N$ assets (see  Lemma \ref{remaoafini}) and also in the large market (see Lemma \ref{AOAINF}).
 \begin{assumption}
\label{AOAfini}
For all $i \geq 1$, $P(\e_i >b_i)>0 \mbox{ and } P(\e_i <b_i)>0.$
\end{assumption}

\begin{lemma}
\label{remaoafini}
Under Assumption \ref{un}, Assumption \ref{AOAfini} implies  AOA($N$) for any $N\geq 1$. Moreover, AOA$(N)$ implies  the so-called \lq{}\lq{}quantitative\rq{}\rq{} no-arbitrage condition:    there exists some $\a_N \in ]0,1[,$ such that
for every $(h_1,\ldots, h_N) \in \R^N$ satisfying $\sum_{i=1}^N h_i^2=1,$ $
P\left(\sum_{i=1}^N h_i(\e_i - b_i)<-{\alpha}_{N}\right)>{\alpha}_{N}.$
\end{lemma}
{\it Proof} Fix some $N \geq 1$ and let $(h_1,\ldots, h_N) \in \R^N$ such that 
$\sum_{i=1}^N h_i(\e_i - b_i) \geq 0$ a.s. We proceed by contradiction. \\
Assume that  $I_N:=\{i \in \{1,\ldots,N\}, \; h_i \neq 0\}\neq \emptyset.$ 
Let $B_i:=\{h_i(\e_i - b_i)<0\}$. Then,  
$\bigcap_{i \in I_N} B_i \subset \left\{\sum_{i=1}^N h_i(\e_i - b_i)<0\right\}.$ 
As the $(\e_i)_{i\geq 1}$ are independent and for $i \in I_N$, 
$P\left(B_i\right)\geq \min\left\{P\left(\{ \e_i - b_i<0\}\right),P\left(\{ \e_i - b_i>0\}\right)\right\}>0,$ we get that 
$P\left(\bigcap_{i \in I_N} B_i\right)= \prod_{i \in I_N}  P\left(B_i\right)>0$, a contradiction.
The proof of the last result is standard (see for example \cite{stettner}) and thus omitted. 
 \qed 

It is well-known that absence of arbitrage in markets with finitely many assets is equivalent to the existence of an equivalent martingale measure,
see \cite{dmw} and the references in \cite{follmer-schied}. In the present setting with infinitely many assets, we need to consider equivalent 
martingale measures having a finite second moment.
Let 
$$
\Mc_ 2:=\left\{Q \sim P: \, {dQ}/{dP} \in L ^2(P), \, E_{Q} (\e_i)=b_i, \, i\geq 1\right\}.
$$

%
\begin{remark}
\label{nulla} {\rm
If $Q\in\mathcal{M}_2$ and if Assumptions \ref{un} and \ref{b} hold true, then for all $h \in \ell_2,$ 
$E_Q\left(V^{0,h}\right)=0$. This is Cauchy-Schwarz inequality,   
see also Lemma 3.4 of \cite{jmaa}.}
\end{remark}


Unfortunately Assumptions \ref{un}, \ref{b} and \ref{AOAfini} are not known to be sufficient to ensure that $\Mc_ 2\neq \emptyset$ 
 (see Proposition 4 of \cite{def}). So we also postulate the following.
\begin{assumption}\label{trois}
We have that 
$
\sup_{i\geq 1} E\left(|\varepsilon_i|^3\right)<\infty.$
\end{assumption}
\begin{remark}
\label{diffmiklos}
We comment on the main differences with \cite{ijtaf,jmaa,def}. First, we use \cite{def}
to show $\mathcal{M}_{2}\neq\emptyset$.{}
This justifies Assumption \ref{trois}.
In \cite{jmaa}, 
both conditions 
\begin{eqnarray}\label{vavelgrof}
\inf_{i\geq 1}P(\varepsilon_i>x)>0\mbox{ and }
\inf_{i\geq 1}P(\varepsilon_i<-x)>0\mbox{ for all }x\geq 0, \\
\label{unint}
\sup_{i\in\mathbb{N}}E\left(\varepsilon_i^2 1_{\{|\varepsilon_i|\geq N\}}\right)\to 0,\ N\to\infty,
\end{eqnarray}
were postulated. It was proved that the set $K^x$ is closed in probability and that for
concave, non-decreasing utility functions
$U:\mathbb{R}\to\mathbb{R}$ 
there exist optimizers. In \cite{ijtaf}, the rather restrictive assumption \eqref{vavelgrof}, which excludes, e.g., the
case where all the $\varepsilon_{i}$ are bounded random variables, was relaxed at the price
of requiring more integrability on the $\varepsilon_{i}$ than \eqref{unint}.
Assumption \ref{AOAfini} was postulated together with 
$
\sup_{i\geq 1} E\left(e^{\gamma |\varepsilon_i|}\right)<\infty,
$ 
for some $\gamma
 > 0.$ This strong moment condition was not justified in the APT 
 problem and in this paper, we manage to use instead the weaker Assumption
 \ref{trois}.
 Moreover, we will be able to prove that 
 $\mathcal{C}^{x}:=K^x -{L}^2_+(P)$ is closed in probability. 
\end{remark}
In Corollary 1 of \cite{def} it is shown that, under
Assumptions  \ref{un}, \ref{AOAfini} and \ref{trois},
\begin{equation}
\label{eqaaa}
\mbox{ AAA } \Longleftrightarrow \mbox{ Assumption}  \, \ref{b}
\Longleftrightarrow \Mc_ 2\neq \emptyset.
\end{equation}

 
Based on \eqref{eqaaa}, one can show that 
AAA implies the classical no-arbitrage condition stated with infinitely many assets.
\begin{lemma}
\label{AOAINF}
Assume that Assumptions \ref{un}, \ref{AOAfini}, \ref{trois} together with AAA holds true. Then,  
$\langle h,\e-b\rangle \geq 0$ a.s. for some $h\in \ell_2$ implies that  $\langle h,\e-b\rangle =0$ a.s.
\end{lemma}
{\it Proof} 
Let $h\in \ell_2$ 
and assume that $\langle h,\e-b\rangle \geq 0$. Fix some $Q \in \Mc_2$ given by \eqref{eqaaa}, then
$E_{Q}(\langle h,\e-b\rangle)=0$  (see Remark \ref{nulla}). Thus
$\langle h,\e-b\rangle=0$ $Q$-a.s. and also $P$-a.s. since $P$ and $Q$ are equivalent.
 \qed 

The following lemma is crucial to prove the closure property of $\mathcal{C}^{x}$ (see Corollary \ref{Cferme}).
\begin{lemma}
\label{miki}
Let Assumptions \ref{un} and \ref{b} hold true and assume, for some $\gamma\geq 2,$ that 
$\sup_{i \geq 1} E|\varepsilon_i|^{\gamma}<\infty$. 
Then, there is a constant $C_{\gamma}$
such that, for all $h\in\ell_2$
$$
E|\langle h,\varepsilon - b\rangle|^{\gamma}\leq 
C_{\gamma}\Vert h\Vert_{\ell_2}^{\gamma} \left(1+ \Vert b\Vert_{\ell_2}^{\gamma}\right).
$$
Moreover, if $\gamma=3,$ for any $c>0,$ $\{|V^{x,h}|^2: \,h \in \ell_2, \, \|h\|_{\ell_2}\leq c \}$ and also $\{|V^{x,h}|: \,h \in \ell_2, \, \|h\|_{\ell_2}\leq c \}$ are  uniformly integrable. 
\end{lemma}


{\it Proof}  Let $h(n):=(h_1,\ldots,h_n,0,0,\ldots)$ and $b(n):=(b_1,\ldots,b_n,0,0,\ldots)$,
for $n \geq 1$. $$E|\langle h(n),\varepsilon -b\rangle|^{\gamma}
= E\left|\sum_{i=1}^{n}h_i(\varepsilon_i-b_i)\right|^{\gamma} \leq 
2^{\gamma-1} E\left|\sum_{i=1}^{n}h_i\varepsilon_i\right|^{\gamma} + 
2^{\gamma-1} E\left|\sum_{i=1}^{n}h_ib_i\right|^{\gamma}.$$ 
The Marcinkiewicz-Zygmund and triangle inequalities imply for some $\bar{C}>0$ 
\begin{eqnarray*}
 E\left|\sum_{i=1}^{n}h_i\varepsilon_i\right|^{\gamma} 
&\leq&  \bar{C} E\left(\left(\sum_{i=1}^{n}h_i^2 \varepsilon_i^2\right)^{\gamma/2}\right)
= \bar{C} \left\|\sum_{i=1}^{n}h_i^2 \varepsilon_i^2 \right\|_{L^{\gamma/2}(P)}^{\gamma/2}\\
&\leq&  \bar{C}\left(\sum_{i=1}^{n}|h_i|^2\left\|\varepsilon_i \right\|^2_{L^{\gamma}(P)}\right)^{\gamma/2}
 \leq 
\bar{C} \left(
 \sup_{i \geq 1} \|\varepsilon_i\|_{L^{\gamma}(P)}^2 \sum_{i=1}^{n}|h_i|^2
\right)^{\gamma/2} \\
&\leq &
\bar{C} 
 \sup_{i \geq 1} E|\varepsilon_i|^{\gamma} \Vert h(n)\Vert_{\ell_2}^{\gamma}. 
\end{eqnarray*}
Thus, $E|\langle h(n),\varepsilon -b\rangle|^{\gamma}
 \leq  C_{\gamma}\Vert h(n)\Vert_{\ell_2}^{\gamma} \left(1+ \Vert b(n)\Vert_{\ell_2}^{\gamma}\right)$
and Fatou's lemma finishes the proof. \qed

For all $x \geq 0$, the set of attainable wealth at time $1$, allowing the possibility of throwing away money,
is ${C}^{x}:=K^x -{L}^2_+(P).$ 
\begin{proposition}
\label{L3}
Let Assumptions \ref{un}, \ref{b}, \ref{AOAfini} and \ref{trois} hold true.
Fix some $z \in \R$ and let $B \in L^2(P)$ such that $B  \notin \mathcal{C}^{z}$.  Then, there exists some $\eta>0$
such that
\begin{align}
\label{Eqvare}
\inf_{h \in \ell_2}  P(z+\langle h, \e-b\rangle < B -\eta)> \eta.
\end{align}
\end{proposition}

{\it Proof} 
Assume that \eqref{Eqvare} is not true.  Then, for all $n \geq 1$, there exists some $h({n}) \in \ell_2$ such that  $P(V_{n} < B -\frac{1}{n})  \leq \frac{1}{n}$, where we have introduced the following notation: $V_{n}:=z+\langle h(n), \e-b\rangle$. Let 
$ 
G_{n}:=\{V_n \geq B-\frac{1}{n} \}
$ and set $\k_{n}:= \left(V_{n} - (B- \frac{1}{n}) \right) 1_{G_n}$. Then, $P\left( |V_{n} -\kappa_{n} -B|> \frac{1}{n}\right) = P\left(\Omega \setminus G_n\right) \leq  \frac{1}{n}$ and  thus, $(V_{n} - \kappa_{n})_{n \geq 1}$ converges to $B$ in probability.

First, we claim that $\sup_n ||h(n)||_{\ell_2}<\infty$. Else,  
$\sup_n ||h(n)||_{\ell_2}=\infty.$ So extracting a subsequence (which we continue to denote by $n$),
we may and will assume that $||h(n)||_{\ell_2}\to\infty$, $n\to\infty$.
Let $\tilde{h}_i(n):=
h_i(n)/||h(n)||_{\ell_2}$ for all $n,i$. Clearly, $\tilde{h}(n)\in \ell_2$
with $||\tilde{h}(n)||_{\ell_2}=1$. Then, 
$$W_n:=
V^{0,\tilde{h}(n)} -\frac{\kappa_n}{||h(n)||_{\ell_2}} \to 0 \mbox{ a.s., } n\to \infty.$$
Let $Q \in \Mc_2$ (which is not empty: see \eqref{eqaaa}). We claim that $
E_{Q} \left(W_n\right) \to 0$. 
By the Cauchy-Schwarz inequality, 
$
\left| E_{Q} \left(W_n\right) \right| \leq  \sqrt{E\left({dQ}/{dP}\right)^2}
\sqrt{E\left(W_n^2\right)}
$ and it remains to show the  uniform integrability of
$W_n^2$, $n\in\mathbb{N}$ under $P$.
\begin{eqnarray*}
\left|W_n\right|^2 & = & \frac{|B-z- n^{-1} |^2}{||h(n)||^2_{\ell_2}}
1_{G_{n}}
+  |V^{0,\tilde{h}(n)} |^2 1_{\Omega \setminus G_{n}}\\
 &  \leq & \frac{|B|^2 +|z |^2 + {n^{-2}}}{||h(n)||^2_{\ell_2}} +  |V^{0,\tilde{h}(n)}|^2 \leq c|B|^2 +|V^{0,\tilde{h}(n)} |^2,
\end{eqnarray*}
for $n$ big enough, with some constant $c$.
Using Assumption \ref{trois} and Lemma \ref{miki}, $|V^{0,\tilde{h}(n)} |^2$, $n\in\mathbb{N}$ for $\|\tilde{h}(n)\|_{\ell_2} \leq 1$ is uniform integrable under $P.$ As $B^2$ is also integrable,  we get that 
$E_{Q} \left(W_n\right)$ goes to $0$.

As $E_{Q} V^{0,\tilde{h}(n)}= 0$ (see Remark \ref{nulla}), we deduce that
${\kappa_n}/{||h(n)||_{\ell_2}}$ goes to zero in $L^1(Q)$ and also $Q$-a.s. (along a subsequence) and, as $Q$ is equivalent to $P$, $P$-a.s. This
implies that
$V^{0,\tilde{h}(n)}$ goes to $0$ $P$-a.s. and in $L^2(P)$  as well (recall that 
the family $|V^{0,\tilde{h}(n)}|^2$, $n\geq 1$ for $\|\tilde{h}(n)\|_{\ell_2} \leq 1$ is uniformly integrable).
But this is
absurd since using the isometry property (see \eqref{isol}), we get that 
$
\|V^{0,\tilde{h}(n)} \|^2_{L^2}    =  \|\tilde h(n)\|^2_{\ell_2}+\sum_{i =1}^{\infty} \tilde h^{2}(n)_ib^{2}_i\geq 1$ for all $n\geq 1$. This contradiction
shows that necessarily $\sup_n ||h(n)||_{\ell_2}<\infty$.

We have concluded that  $\sup_n ||h(n)||_{\ell_2}<\infty$. 
Since $\ell_2$ has the Banach-Saks property, there exists a
subsequence $(n_k)_{k\geq 1}$ and, some $h^*\in \ell_2,$ such that
$$\widehat{h}(N):=\frac{1}{N}\sum_{k=1}^N h (n_k), \;\;\;
\left\Vert\widehat{h}(N)-h^*\right\Vert_{\ell_2}^2 \to 0,\ N\to\infty.
$$ 
Hence, using \eqref{isol}, 
$
E\left(\left(V^{z,\widehat h(N)}-V^{z, h^*}\right)^2\right)
 \leq  (1+\|b\|_{\ell_2}^2) \| \widehat{h}(N) -h^*\|^2_{\ell_2},$ 
which tends to zero as $N\to\infty$. So $V^{z,\widehat h(N)}\to V^{z,h^*}$ a.s. as well. Then,
\begin{eqnarray*}
V^{z,\widehat h(N)}-\frac{1}{N}\sum_{k=1}^N \kappa_{n_k}=\frac{1}{N}\sum_{k=1}^N\left(
V^{z,{h}(n_k)} - \kappa_{n_k}\right) \to B,\ N\to\infty,
\end{eqnarray*}
in probability, and also a.s., for a subsequence for which we keep the same notation. Thus, $\frac{1}{N}\sum_{k=1}^N \kappa_{n_k}$ converges a.s. and  
$B \in \mathcal{C}^{z}$, a contradiction. \qed

\begin{corollary}
\label{Cferme}
Let Assumptions \ref{un}, \ref{b}, \ref{AOAfini} and \ref{trois} hold true and  fix some $z \in \R$. Then, 
 $\mathcal{C}^{z}$ is closed in probability.
\end{corollary}
{\it Proof} 
Assume that $ \mathcal{C}^{z}$ is not closed in probability. 
Then, one can find some $h(n) \in  \ell_2$ and $\kappa_n \in {L}^2_+(P)$ 
such that $\theta_n:=z+\langle h(n), \e-b\rangle-\kappa_n\in \mathcal{C}^{z}$ converges in probability to some $\theta^* \notin \mathcal{C}^{z}$. Then, for any $\eta>0$,
\begin{eqnarray*}
\inf_{h \in \ell_2}  P(z+\langle h, \e-b\rangle < \theta^*  -\eta)\leq   P(z+\langle h(n), \e-b\rangle-\kappa_n < \theta^*  -\eta) \to 0,
\end{eqnarray*}
when $\eta$ goes to zero. This contradicts \eqref{Eqvare}, showing closedness of $C^z$. \qed

We now provide a quantitative version of the no-arbitrage condition 
(see Assumption \ref{AOAfini}).

\begin{proposition}
\label{aoaquant}
Let Assumptions \ref{un}, \ref{b}, \ref{AOAfini} and \ref{trois} hold true.
Then, there exists $\a>0$, such that for all $h \in \ell_2$ with $\|h\|_{\ell_2}=1,$ 
 $P(\langle h,\e-b\rangle<-\a) > \a$ holds.
\end{proposition}

{}

{}
{\it Proof}  We argue by contradiction.
Assume that for all $n\geq 1$, there exist $h(n)$ with $\|h(n)\|_{\ell_2}=1$ and 
$P\left(\langle h(n),\e-b \rangle  <-{1}/{n}\right) \leq {1}/{n}$. \\Clearly, $\langle h(n),\e-b\rangle_{-}\to 0$ in probability as $n\to\infty$. 
Let $Q \in \Mc_2$ (see \eqref{eqaaa}). We claim that $E_{Q}(\langle h(n),\e-b\rangle_{-})\to 0$. Using Cauchy-Schwarz inequality 
\begin{eqnarray*}
E_{Q}\left(\langle h(n),\e-b \rangle_{-}\right) 
& \leq &  \|dQ/dP\|_{L^2(P)} \left(E\left(\langle h(n),\e-b\rangle_-^{2}\right)	\right)^{1/2},
\end{eqnarray*}
and it remains to show uniform integrability of $\langle h(n),\e-b\rangle_-^{2},$ $n \in \N$ under $P$. This follows from 
 $\langle h(n),\e-b\rangle_-^{2} \leq |V^{0,h(n)}|^2$,  Assumption \ref{trois} and Lemma \ref{miki}. 
So $E_{Q}(\langle h(n),\e-b\rangle_{-})\to 0$ but, since
$E_{Q}(\langle h(n),\e-b\rangle)=0$ by Remark \ref{nulla}, we also get that $E(\langle h(n),\e-b\rangle_{+})\to 0$. 
It follows that
$E_{Q}(|\langle h(n),\e-b\rangle|)\to 0,$
hence $\langle h(n),\e-b\rangle$ goes to zero $Q$-a.s. (along a subsequence) and, as $Q$ is equivalent to $P$, $P$-a.s.  Using again that $|\langle h(n),\e-b\rangle|^{2}$, $n\in\mathbb{N}$ is 
uniformly $P$-integrable, we get $E(|\langle h(n),\e-b\rangle|^{2})\to 0.$ But this
contradicts the fact that $E(|\langle h(n),\e-b\rangle|^{2})=\Vert h(n)\Vert_{\ell_2}^{2}+ \sum_{i=1}^{\infty}h^{2}_{i}(n)b_{i}^{2}
\geq 1$ (see \eqref{isol}).\qed

The following lemma proves that, under the no-arbitrage condition (see Assumption \ref{AOAfini}),  
any strategy with a non-negative final wealth is bounded. 

\begin{lemma}
\label{hborne}
Let Assumptions \ref{un}, \ref{b},  \ref{AOAfini} and \ref{trois} hold true.
Let $y\in \R$ and $h \in \ell_2$ such that $y+\langle h, \e-b\rangle  \geq 0$. Then, $\|h\|_{\ell_2}\leq {|y|}/{\a},$
see Proposition \ref{aoaquant} for $\a$. 
\end{lemma}

{\it Proof} 
On $\{\langle h,\e-b\rangle<-\a ||h||_{\ell_{2}} \}$, which is of positive measure by Proposition \ref{aoaquant}, 
$|y|-\a ||h||_{\ell_{2}}  >y+\langle h, \e-b\rangle \geq 0$ and  $\|h\|_{\ell_2}\leq {|y|}/{\a}$ follows.  \qed

\section{Superreplication Price}\label{supe}

Let $G\in L^{0}$
be a random variable, which will be interpreted as the payoff
of some derivative security at time $T$.
The superreplication price $\pi(G)$ is
the minimal initial wealth needed for hedging $G$ without risk. For all $x\in \R$, let 
$$\mathcal{A}(G,x):=\left\{h \in \ell_2:\; V^{x,h} \geq G\; \mbox{a.s.} \right\} \mbox{and } 
\pi(G):=\inf\{z\in\mathbb{R}:\ \mathcal{A}(G,z)\neq \emptyset\}, 
$$
where  $\pi(G)=+ \infty$ if $\mathcal{A}(G,z)=\emptyset$ for every $z$. The so-called dual representation of the superreplication price (see Theorem
 \ref{thmsur} below) in terms of supremum over the different risk-neutral probability measures has a long history: see \cite{cvitanic-karatzas} 
and also the textbook \cite{follmer-schied} 
for more details about this preference-free price.
 \begin{lemma}
 \label{lemsur}
Let Assumptions \ref{un}, \ref{b}, \ref{AOAfini}  and  \ref{trois} holds true. Then, $\pi(G)>-\infty$ and $\mathcal{A}(G,\pi(G))\neq \emptyset.$ 
 \end{lemma}
 {\it Proof} 
 Assume that $\pi(G)=-\infty$. Then, for all $n \geq 1$, there exists
$h_n \in \ell_2$ such that $-n+\langle h_n,\e-b\rangle\geq G$ a.s. Thus, $\langle h_n,\e-b\rangle\geq G+n\geq (G+n) \wedge 1$ a.s. It follows that
$(G+n) \wedge 1 \in C^0,$ 
which is closed in probability (see Corollary \ref{Cferme}). Thus, $1 \in C^0$, i.e.,  $\langle h,\e-b\rangle\geq 1$ a.s. for some $h \in \ell_2$, which contradicts AAA (or Assumption \ref{b}, see \eqref{eqaaa}), see Lemma \ref{AOAINF}. So $\pi(G)>-\infty$.\\
If $\pi(G)=+\infty$, the second claim is trivial. So, assume that,
$\pi(G)<\infty$. Then, for all $n \geq 1$, there exists
$h_n \in \ell_2$ such that $\pi(G) +1/n+\langle h_n,\e-b\rangle\geq G \mbox{ a.s.}$  It follows that
$G- \pi(G) -1/n\in C^0.$ Thus, as $C^0$ is closed, $G- \pi(G)\in C^0$. \qed

We are now in position to prove our duality result.
 \begin{theorem}
 \label{thmsur} 
Let Assumptions \ref{un}, \ref{b}, \ref{AOAfini}  and  \ref{trois} hold true and let $G \in L^2(P)$. Then,  
$\pi(G)=\sup_{Q \in \Mc_2}E_{Q}(G)$.
 \end{theorem}
{\it Proof} 
Let $s:=\sup_{Q \in \Mc_2}E_{Q}(G)$.
Let $x$ be such that there exists
$h \in \ell_2$ verifying $x +\langle h,\e-b\rangle\geq G$ a.s.
Fix $Q \in \Mc_2$ (see \eqref{eqaaa}). As $G \in L^2(P)$, $E_Q(G)$ is well-defined 
by the Cauchy-Schwarz inequality. Using Remark \ref{nulla}, we get that $E_Q(x +\langle h,\e-b\rangle)=x$. Thus, $x \geq  E_Q(G)$ and 
$\pi(G) \geq s$ follows. For the other inequality, it is enough to prove that
$G- s\in C^0$. Indeed, this will imply that
there exists $h \in \ell_2$ such that $s+\langle h,\e-b\rangle\geq G$ a.s., which shows, by definition of $\pi(G)$, that
$s\geq \pi(G)$. 
Assume this is not true. Then, $\{G- s\} \notin C^0\cap L^2(P).$
As $C^0$ is closed in probability (see Corollary  \ref{Cferme}), we can apply classical  
Hahn-Banach argument (see, e.g., \cite{follmer-schied}) to find some $Q \in \Mc_2$ 
such that $E_{Q}(G) > s$. \qed

%

 \begin{remark}
 \label{surgrandpetit}
 One may wonder whether $\pi_n(G)$, the superreplication price of $G$ in the small 
 market with $n$ random sources $(\e_i)_{1 \leq i \leq n}$, converges to $\pi(G)$, 
 the superreplication price of $G$ in the large market. The answer is \emph{no} in general: 
 let $\varepsilon_i$, $i\in\mathbb{N}$ be standard Gaussian random variables, let $b_i=0$
for all $i\in\mathbb{N}$ and define
$G:=\sum_{i=1}^{\infty} i^{-1}\varepsilon_i$. There exists no $x,h_1,\ldots,h_n$ with
$ 
x+\sum_{j=1}^n h_j\varepsilon_j\geq G,
$ 
since this would mean that
$ 
\sum_{j=1}^n (h_j-j^{-1})\varepsilon_j-\sum_{j \geq n+1} j^{-1}\varepsilon_j\geq -x, 
$ 
where the left-hand side is a Gaussian random variable with non-zero variance. It follows that $\pi_n(G)=\infty$
while $\pi(G)=0$, trivially.
\end{remark}

\section{Utility Maximization}\label{negu}

We follow the traditional viewpoint of \cite{von} and model economic agents' preferences by
some concave
strictly increasing differentiable utility function  denoted by $U:]0,\infty[\to\mathbb{R}$. Note that we extend  $U$ to $[0,\infty[$ by (right)-continuity ($U(0)$ may
be $-\infty$). We also set $U(x)=-\infty$ for $x\in ]-\infty,0[$.
For a contingent claim $G \in  L^0$  and $x \in \mathbb{R} $, we define 
$\Phi(U,G,x):=\left\{ h \in \ell_2,\; E U^{+}(V^{x,h}-G)<+\infty \right\},$  the set of strategies, where the expectation is well-defined. 
Then, we  set $\mathcal{A}(U,G,x):= \Phi(U,G,x) \cap \mathcal{A}(G,x).$ 
Note that even for $x  \geq \pi(G)$, $\mathcal{A}(U,G,x)$ might be  empty. Indeed, from  Lemma \ref{lemsur}, we know that there exists some  $h \in \mathcal{A}(G,x)$,
but $h$ might not belong to $\Phi(U,G,x)$. But this holds true under appropriate assumptions, as proved in the lemma below.
\begin{lemma}
\label{toutva} Let Assumptions \ref{un}, \ref{b},  \ref{AOAfini}  and \ref{trois}  hold true. Assume that
$G \geq 0$ a.s. and $U(x_0)=0$, $U'(x_0)=1$, for some $x_0 \geq 0$. Then, 
$\mathcal{A}(G,x)=\mathcal{A}(U,G,x)$ for all $x \in \R$.
\end{lemma}
{\it Proof} 
As $U$ is concave, increasing and differentiable with $U(x_0)=0,$ 
$U\rq{}(x_0)=1$, we can bound it from above 
by its first order Taylor approximation, for all $x\in ]0,\infty[$, as follows:
\begin{eqnarray*}
U(x) & \leq & U(\max(x_0,x))\leq U(x_0)+ \max(x-x_0,0) U\rq{}(x_0)
 \leq |x-x_0|\leq |x|,
\end{eqnarray*}
since $x_0 \geq 0.$ If $x <\pi(G)$ then $\mathcal{A}(G,x)=\emptyset$ and $\mathcal{A}(G,x)=\mathcal{A}(U,G,x)=\emptyset$.
Let $x \geq  \pi(G)$. Then, by Lemma \ref{lemsur}, $\mathcal{A}(G,x)\neq \emptyset.$
Let $h \in \mathcal{A}(G,x)$. Then, $V^{x,h} \geq G \geq 0\; \mbox{a.s.}$ and $h \in \mathcal{A}(0,x)$. Let $A:=\{x+\langle h, \e-b\rangle \geq x_0\}.$ 
\begin{eqnarray}
\nonumber
U^+(x+\langle h, \e-b\rangle-G)
& \leq &   
U^+(x+\langle h, \e-b\rangle) 1_{A} +  U^+(x_0) 1_{\Omega \setminus A} \\
\label{bolus}
& = &  U(x+\langle h, \e-b\rangle) 1_{A}  \leq  |x+<h, \e-b>|. 
\end{eqnarray}
Using \eqref{isol}, the Cauchy-Schwarz inequality and  Lemma \ref{hborne}, we get that
\begin{eqnarray}
\nonumber
EU^+(x+\langle h, \e-b\rangle-G) & \leq &  |x| +
\sqrt{E\left(\langle h,\varepsilon -b\rangle^2\right)}  
 \leq  |x| + \| h\|_{\ell_2}\sqrt{1+\|b\|^2_{\ell_2}}\\
\label{ilfaitbeau}
& \leq &|x|+ \frac{|x|}{\a}\sqrt{1+\|b\|^2_{\ell_2}}
<+\infty.
\end{eqnarray}
\qed

We now define  the supremum of the expected utility at the terminal date when delivering the claim $G$, starting
from initial wealth $x \in \R$ :
\begin{eqnarray}
\label{gnon}
u(G,x):=\sup_{h \in\mathcal{A}(U,G,x)}EU(V^{x,h}-G),
\end{eqnarray}
where $u(G,x)= -\infty$, if $\mathcal{A}(U,G,x)= \emptyset$.
The following result establishes that there exists an optimal investment for
the investor we are considering.

\begin{theorem}\label{csonti}
Let Assumptions \ref{un}, \ref{b},  \ref{AOAfini} and \ref{trois} hold true. Let $G\geq 0$ and $x \in \R$ such that  
$x \geq \pi(G)$. Then, there exists
$h^*\in\mathcal{A}(U,G,x)$ such that $$
u(G,x)=EU(V^{x,h^*}-G).
$$
\end{theorem}
{\it Proof} 
If $U$ is constant, there is nothing to prove. Else, there exists $x_0>0$ such that $U\rq{}(x_0)>0$. Replacing $U$ 
 by $({U}-{U(x_0)})/{U\rq{}(x_0)}$, we may and will suppose that $
U(x_0)=0$ and  $U'(x_0)=1.$ Note that $\pi(G) \geq 0,$ as $G \geq 0$ a.s. (see Theorem \ref{thmsur}). Let $h_n\in\mathcal{A}(G,x)=\mathcal{A}(U,G,x)$ (see Lemmata \ref{lemsur} and  \ref{toutva}) be a sequence such that $$
EU(V^{x,h_n}-G)\uparrow u(G,x),\ n\to\infty.
$$
By Lemma \ref{hborne},
$\sup_{n\in\mathbb{N}} \|h_n\|_{\ell_2}\leq {x}/{\a}<\infty.$
Hence, as $\ell_2$ has the Banach-Saks Property, there exists a subsequence $(n_k)_{k\geq 1}$ and some $h^*\in \ell_2$ such that for $\tilde{h}_n:=\frac{1}{n}\sum_{k=1}^n h_{n_k},$  $\|\tilde{h}_n-h^*\|_{\ell_2}\to 0, \,n\to\infty.$ 
 Note that $\tilde{h}_n\in\mathcal{A}(G,x)$ and $\sup_{n\in\mathbb{N}} \|\tilde h_n\|_{\ell_2}\leq {x}/{\a}<\infty.$
Using \eqref{isol}, we get that 
$$
E\langle \tilde{h}_n-h^*,\varepsilon-b\rangle^2
 \leq  \| \tilde{h}_n-h^*\|^2_{\ell_2}(1+\|b\|^2_{\ell_2})\to 0, \,n\to\infty,$$ 
In particular, $\langle \tilde{h}_n-h^*,\varepsilon-b\rangle\to 0$, $n\to\infty$
in probability. Hence,  we also get that $U(V^{x,\tilde{h}_n}-G)\to U(V^{x,h^*}-G)$ in probability, by continuity (right-continuity in $0$) 
of $U$ on $[0,\infty[$.
We also have (up to a subsequence) that $V^{x,\tilde{h}_n}-G\to V^{x,h^*}-G$ a.s.  and thus, $h^* \in \mathcal{A}(G,x)$.
Now, using \eqref{bolus}, we have that 
$
U^+(V^{x,\tilde{h}_n}-G)\leq
|V^{x,\tilde{h}_n}|.
$  
So Assumption \ref{trois} and Lemma \ref{miki} imply that  
$\{U^+(V^{x,\tilde{h}_n}-G): \, h_n\in \ell_2, \, \|h_n\|_{\ell_2} \leq {x}/{\a}\}$ is uniformly integrable and 
$$\lim_{n\to\infty}E\left(U^+(V^{x,\tilde{h}_n}-G)\right)=
E\left(U^+(V^{x,{h}^*}-G)\right).$$
Then, 
$E\left(-U^-(V^{x,h^*}-G)\right)\geq \limsup_{n\to\infty}E\left(-U^-(V^{x,\tilde{h}_n}-G)\right),$ by Fatou's lemma. As 
by  concavity of $U,$
$$
U(V^{x,\tilde{h}_n}-G)= U\left(\frac{1}{n}\sum_{k=1}^n  (V^{x,{h}_{n_k}}-G)\right)\geq
\frac{1}{n}\sum_{k=1}^n U\left( V^{x,{h}_{n_k}}-G\right
),$$ we get that
$$ 
EU(V^{x,h^*}-G)\geq \limsup_{n\to\infty}EU(V^{x,\tilde{h}_n}-G) \geq  u(G,x). 
$$
The proof is finished since $h^*\in\mathcal{A}(G,x)=\mathcal{A}(U,G,x)$ (see Lemma \ref{toutva}).
\qed



\section{Convergence of the Reservation Price to the Superreplication Price}\label{cove}

We go on incorporating a sequence of agents in our model. 
\begin{assumption}\label{U} Suppose that
$U_n:]0,\infty[\to\mathbb{R},\ n\in\mathbb{N}$ is a sequence of concave
strictly increasing twice continuously differentiable functions such that
$$
\forall x\in ]0,\infty[\quad
r_n(x):=-\frac{U_n''(x)}{U_n'(x)}\to\infty,\ n\to\infty.
$$
\end{assumption}
Again we extend each $U_n$ to $[0,\infty[$ by (right)-continuity, and set $U_n(x)=-\infty$ for  $x\in ]-\infty,0[$. 
We define the value functions $u_n(G,x)$  for our sequence of utility functions $(U_n)_{n \geq 1}$ changing $U$ by  $U_n$ in  \eqref{gnon}. \\
Assumption \ref{U} says that the sequence of agents we consider
have asymptotically infinite aversion towards risk. Indeed, \cite{pratt} shows  that an investor $n$ has greater absolute risk-aversion than investor $m$ (i.e.,
$r_n(x) > r_m(x)$ for all $x$) if and only if investor $n$ is more risk averse than $m$ (i.e., the amount of cash for which she would exchange the risk 
is smaller for $n$ than for $m$).  \\
The utility indifference (or reservation) price $p_n(G,x)$, introduced by \cite{hodges-neuberger}, is
$$ 
p_n(G,x):=\inf\{z\in\mathbb{R}: u_n(G,x+z)\geq u_n(0,x)\}.
$$
Intuitively, it seems reasonable that under Assumption \ref{U}
the utility prices $p_n(G,x)$ tend to $\pi(G)$ and this was proved for finitely many assets in
\cite{laurence-miklos}. Now, we treat the case of APT.

\begin{theorem}\label{main} Assume that
Assumptions \ref{un}, \ref{b}, \ref{AOAfini}, \ref{trois} and \ref{U}  hold true. Suppose that $x>0$ and $G \in L^2_+(P)$. Then,  the utility indifference prices $p_n(G,x)$ are
well-defined and converge
to $\pi(G)$ as $n\to\infty$.
\end{theorem}
{\it Proof} 
Applying affine transformations to each $U_{n},$ we may and will assume that 
$U_n(x)=0$ and  $U_n'(x)=1$ for all $n\in\mathbb{N}$.


If $\pi(G)=+\infty$ then for all $z \in \mathbb{R}$, $n \geq 1$, 
$\emptyset=\mathcal{A}(G,z)=\mathcal{A}({U}_{n},G,z)$ and ${u}_{n}(G,x+z)=-\infty$. 
But $u_n(0,x) \geq EU_n(x)=0$. Thus,  ${p}_{n}(G,x)=+\infty$ for all $n \geq 1$ and the claim is proved. \\
Assume now that $\pi(G)<\infty$. Just like in the proof of Theorem 3 in \cite{laurence-miklos}, $p_n(G,x)\leq \pi(G).$ 
So,  it remains to show that $\liminf_{n \to \infty} p_{n}(G,x) \geq \pi(G)$. If this is not the case, we can find a subsequence (still denoted by $n$)  and some $\eta >0$ such that
$p_{n}(G,x) \leq  \pi(G)-\eta$ for all $n \geq 1$. We may and will assume that $ x \geq \eta$.
By definition of $p_{n}(G,x),$ we have that 
$$u_{n}(G,x + \pi(G)-\eta) \geq u_{n}(0,x).$$
Let $y:=x + \pi(G)-\eta<x + \pi(G)$. If we prove that 
 $\lim_{n \to +\infty} u_{n}(G,y) = -\infty,$ $\liminf_{n \to +\infty} u_{n}(0,x) \geq \liminf_{n \to +\infty} U_{n}(x) \geq 0$ 
will provide  a contradiction.  \\
First, remark that  $x+ G \notin \mathcal{C}^{y}$.  
Applying  Proposition \ref{L3},
we get some $\gamma>0$ such that $\inf_{h \in \ell_2}  P(A_{h})> \gamma$, where $A_{h}:=\{
y+\langle h, \e-b\rangle <x+ G -\gamma \}.$
Note that we can always assume that $x \geq \gamma$. As $ y\geq \pi(G) \geq 0$,  Lemmata \ref{lemsur} and \ref{toutva} imply that $\mathcal{A}(U_n,G,y) \neq \emptyset.$ Hence, for all $h \in \mathcal{A}(U_n,G,y)$, we get that
\begin{eqnarray*}
\nonumber  EU_n(y+\langle h, \e-b\rangle-G)
& \leq &   E1_{A_{h}}U_n(x-\gamma)+
E1_{\Omega \setminus A_{h}}U^+_n( y+ h \langle \e-b \rangle  ) \\
\label{estimation} & \leq &  \gamma U_n(x-\gamma)+ EU^+_n(y+\langle h, \e-b\rangle).
\end{eqnarray*}
Using  \eqref{ilfaitbeau}, 
$
u_n(G,y)\leq \gamma U_n(x-\gamma)+ y+\frac{y}{ \a}\sqrt{1+\|b\|^2_{\ell_2}}$ goes to
$-\infty$ 
when $n$ goes to infinity, by Lemma 4 of \cite{laurence-miklos}.
\qed

\section{Conclusions}
\label{conclu}
The current paper, just like \cite{ijtaf,jmaa,def},
is based on techniques that are at the intersection of probability
and functional analysis. These permit to state a dual representation for the super-replication cost, to prove existence in the problem of maximization of expected utility and to show the convergence of the reservation prices to the super-replication cost in markets with infinitely
many assets, which form an important model class of financial mathematics, pertinent
to, e.g., bond markets. In future work, our approach is hoped to be extended to other
infinite market models (e.g., complete ones, where $\varepsilon_{i}$ are not independent
but form a complete orthonormal system) so as to gain further insight about how these
complex systems operate.

\medskip

\noindent\textbf{Acknowledgments}

\smallskip

\noindent M.R. was supported by NKFIH grant KH 126505 and by grant LP 2015-6
of the Hungarian Academy of Sciences.


\end{document}